# Application of TAM model to the use of information technology

**Harryanto [1] *, Muchriana Muchran [2], Ansari Saleh Ahmar [3]**

[1] *Department of Accounting, Hasanuddin University, Makassar, Indonesia*
[2] *Department of Accounting, Muhammadiyah University of Makassar, Makassar, Indonesia*
[3] *Department of Statistics, Universitas Negeri Makassar, Makassar, Indonesia*
*\*Corresponding author E-mail: endhyharryanto@gmail.com*

**Abstract**

The purpose of this research is to see the application of modified TAM model by entering the experiential variable as a moderation variable to see one's intention in the use of technology especially internet banking. Data obtained through the distribution of questionnaires to cus-tomers. The study population is bank customers registered as users of internet banking services. The sample selection used a simple random sampling technique. Hypothesis testing using Partial Least Square (PLS) method through AMOS program. The results showed that the proposed five hypothesis, two significant and three insignificant. Perceived ease of use is significantly related to perceived usefulness. Per-ceived usefulness is not significantly related to intention to use. Perceived ease of use is significantly related to intention to use moderated by experience and not significantly correlated with intention to use moderated by the Experience.

## 1. Introduction

The current global era, information systems are important in helping organizations run their activities. Without a good information system, it will be very difficult, given that the current global flow has become one of the main needs of the activities in an organization. An information system becomes indispensable as it can assist in carrying out the activities undertaken within the organization. Information systems today have become a major requirement in the running of the organization.

In accepting a new technology system, not everyone will can be well understood [1], [2]. Therefore, it is important to assess or measure the level of acceptance and understanding recipients and users of information technology by measurement behavior of the user.

Some researchers compare three theories of behavior, namely: technology acceptance model (TAM), Theory of Reasoned Action (TRA) and theory of planned behavior (TPB). Davis et.al. found that TAM was better at explaining the willingness to accept technology than TRA [3]. Mathieson in his research concludes that TAM and TPB explain behavioral intentions well, but TAM explains the attitude (attitude) better than TPB and TAM model is simpler than TPB [4]. It is common not to blame one model better than the other.

Chau and Hu also compared these three models and the results of a study showing better TAMs explaining the use of technologies disclosed by TPB [5]. Technology Acceptance Model was first introduced by Davis [6]. This theory was developed from Theory of Reasoned Action or TRA by Ajzen and Fishbein [7]. TAM adds two main constructs to the TRA model are perceived usefulness and perceptions of ease of use. Both of these factors affect the intention to use information technology (intention to use) before until the actual function is created in everyday life (actual use).

Research by Gardner and Amoroso [8] developed TAM by gouging out different variables for use by internet users. The four external variables are gender, experience (experience), complexity (complexity), and volunteerism (volunteerism). The results show that external variables influence perception of utility, freedom perception and behavioral intention.

In this study, researchers construct the TAM model by incorporating experience variables because it assumes that experience is also a determinant of behavior. In particular, experiences from the past will help to realize the intent. This is what causes calculation in the formation of intentions [7]. Gardner and Amoroso [8] develop TAM by charging other variables to use the internet. External variables are gender, experience, complexity, and volunteerism. The research results are external variables that affect both usability, freedom perspectives and behavioral intentions. Taylor and Todd [9] found significant differences between experienced system users and those not experienced in using the system. Lessons and information about people with different levels of experience from people with low levels of experience [10].

The purpose of this research is to see the application of modified TAM model by incorporating the moderation variable to see one's intention in the use of technology especially internet banking.

## 2. Theory based

### 2.1. Technology acceptance model (TAM)

The Technology Acceptance Model (TAM) is a framework developed by Fred D. Davis in 1986. Davis's model in the adaptation of Theory Reasoned Action which assumes that one adopts a technology is generally determined by the cognitive process and aims to satisfy the wearer or maximize the usefulness of the technology. TAM is used to examine and measure factors that influence decisions whether one accepts or rejects the information technology. The TAM model is developed from psychological theory that explains that computer user behaviour is based on belief, attitude, intention, and user behaviour relationship. The purpose of this model is to explain the main factors of user behaviour toward acceptance technology users. In more detail explain the acceptance of IT with certain dimensions that can affect the acceptance of IT by the user (user).

### 3.2. Perceived ease of use

According to Davis [1] Perceived ease of use is defined as a measure in which a person believes that a computer can be easily understood and used. Meanwhile, according to Jogiyanto [11] perception of ease is defined as the extent to which a person believes that by using technology will be free of a business so that if person believe that the information system easy to use then he will use it and vice versa.

### 2.3. Perceived usefulness

According to Davis [3], perceived usefulness is defined as a measure by which the use of technology is believed to provide benefits to the person using it and the perception of usefulness as the subjective ability of future users where using a specific application system will improve performance in the organizational context. Usability perception is a level where one believes that the use of a particular technology will provide benefits or provide a positive impact that will be obtained when using the technology

### 2.4. Behavioural intention

Behavioural intention to use is tend behaviour of a person in doing technology. Interest in behaviour can be seen from the level of technology use so it can be predicted from the attitude and attention. The motivation to keep using such technology, as well as the desire to motivate other users

### 2.5. Model research

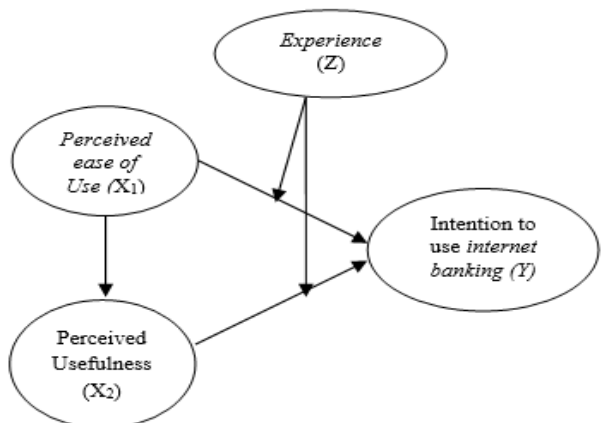

**Fig. 1:** Research Model.

It can explicitly be restated in the form of a linear function to be estimated by simultaneous linear regression as follows

$Y = \beta_0 + \beta_1 X_1 + \beta_2 Y_2 + \beta_3 Z_1 + \beta_4 Z_2 + e$

Where:
$X_1$ = perception of ease of use (perception of convenience of using)
$X_2$ = perceived benefits (Benefit Perception)
$Y$ = Intention to use (Adoption intent)
$Z$ = Experience.

### 2.6. Research method

This research is quantitative approach, it is said quantitative approach because the approach used in the proposed research, process, hypothesis, down the field, data analysis and data conclusions up to the writing using measurement aspects, calculations, formulas and numerical data certainty.

In TAM research with complicated models, basically can be analysed by using regression analysis and path analysis, but in practice will be inefficient because each sub-structure must be analysed one by one, then merged into a whole model. Such complex problems can be analysed using Structural Equation Modelling (SEM) analysis.

In this research using structural equation modeling (SEM) method and analysis tool used in this method is Partial Least Square (PLS) by using AMOS program.

## 3. Results

Respondents who were samples in this study were customers who registered as internet banking users. The number of questionnaires distributed as many as 100 sheets and which returned as many as 95 sheets.

**Table 1:** Research Result

| Directions Influence between Variables | Path coefficient | t-Statistic | Probability | Information |
|---|---|---|---|---|
| $X_1 \rightarrow X_2$ | 0.596 | 3.070 | 0.002 | Accepted |
| $X_1 \rightarrow Y$ | 0.495 | 1.827 | 0.068 | Rejected |
| $X_2 \rightarrow Y$ | -0.008 | -0.039 | 0.969 | Rejected |
| $X_1 Z \rightarrow Y$ | 0.429 | 6.618 | 0.000 | Accepted |
| $X_2 Z \rightarrow Y_3$ | 0.035 | 0.972 | 0.331 | Rejected |

*) Significant on $\alpha = 5\%$; N=95.

## 4. Discussion

### 4.1. The effect of perceived ease of use to perceived usefulness

Perceived ease of use (PEU) is the perceived ease of use of internet banking perceived by users. These variables are measured through ease of learning, clarity and easy to understand, ease of skill, and ease of use.

Perceived usefulness (PU) is the perceived benefit of internet banking users perceived by users. This variable is measured through performance improvement, increased effectiveness, increased productivity, and benefits in transacting.

The results show that perceived ease of use has a significant influence on perceived usefulness. This means that when customers feel the convenience of Internet Banking, customers will also benefit in using Internet Banking.

Perceived usefulness is a belief about the decision-making process. Thus if one feels confident that the information system is easy to use then he will use it. Conversely, if one believes that the information system is not easy to use then he will not use it [9].

The significant relationship between perception of ease of use and perception of benefits is in accordance with TAM Theory which states that a customer will feel that using internet banking can be useful when the use is also easy for them. The easier the operation of an internet banking then this will attract customers to continue to use internet banking. Through continuous use, customers are increasingly able to feel the usefulness of internet banking.

The results of this study support the results of research conducted by Wang, et. al. [12], Al-Somali, et al [13], and Davis, et. Al. [3] which states there is a significant relationship between perceived ease of use and perception of benefits.

Davis et. al. [6] says that the perception of ease of use (PEU) is the degree to which a person believes that using a particular system does not require any effort (free of effort). Wang, et al. [12] says that a customer will feel that using internet banking can be useful when the use is also easy for them. The easier the operation of an internet banking then this will attract customers to continue to use internet banking. Al-Somali, et al. [13] says that with continuous usage, customers are increasingly able to feel the usefulness of internet banking.

### 4.2. The effect of perceived ease of use on intention to use

Intention to use in the context of internet banking is the intention of the user in accepting whether to use internet banking services or not use it. This variable is measured through the intention of continuing to use internet baking, using intentions when having access and suggesting to others to use.

The results show that perceived ease of use has no significant relationship to intention to use. This means that the intention of someone in using internet banking is not related to perceived convenience, or in other words, even if the customer feel the ease in internet banking service is not necessarily the customer will use the internet banking services. This result is not in accordance with the initial hypothesis that there is a significant relationship between perceived ease of use of intention to use.

In the theory of motivation is known the theory of inrinsic and extrinsic motivation. TAM's own construct is the ease of use of perception referred to as intrinsic factor. Intrinsic motivation is when the nature of the work itself that makes a person motivated, the person gets satisfaction by doing the job not because of other stimuli such as status or money or it could be said to do a hobby. Characteristics of respondents in this study indicate that users have a high level of education and income. This indicates that respondents have often used technology, either while still sitting on a lecture bench or may have become a hobby, which is currently increasingly mushrooming social networking sites, which is a major factor in increasing Internet users. With frequent interaction with computers and the internet, the use of internet banking is not difficult to learn based on the experience that has been owned, so in receiving internet banking, they do not see from the convenience anymore, but from other factors.

Judging from the aspect of motivation, it turns out TAM still has a weakness in measuring the motive of acceptance of real information systems. If scrutinized, TAM research only focuses on the question of the utility and effectiveness of a system for task completion. TAM has not considered other variables such as the benefits of information systems for the enjoyment or likes of system users. On the weakness of TAM, Davis, Bagozzi, and Warshaw developed TAM [14], and the result proves that indeed the acceptance of an information system is also determined by another intrinsic motivation factor called perceived enjoyment.

These results support the research of Chau and hu [3], which examines the application of the Technology Acceptance Model (TAM) in explaining the physician's decision to accept telemedicine technology in a health context. The results showed that perceived ease of use constructs did not significantly affect intention to use intention to use. Professional physicians can show considerable differences in general competence, adaptability to new technologies, intellectual and cognitive capacities, and the nature of their work. This is because doctors do not want to spend time learning new technology, even if it's very easy to use. This is especially true when adoption and use of technology may interfere with the routine of their traditional practices.

The results of this study rejected the Pikkaranien [14] study. Research conducted in Finland examines the variables that may affect the use of internet banking in such research such as benefits, ease of use, enjoyment, online banking information, security and privacy, and the quality of internet connection. The results show that ease of use has a significant effect on the intentions of online banking use. Although these results have a smaller value when compared with the benefits.

### 4.3. The effect of perceived ease of use on intention to use which is moderated by experience

The results show that Perceived ease of use has a significant relationship to Intention to use that is moderated by Experience. This means that ease of perceived customers in using internet banking will affect the intention in using internet banking when the customer already has experience with information technology. These results are consistent with the initial hypothesis that there is a significant relationship between trusts against perceived ease of use.

Igbaria et al. [15] suggests that the experience of using technology will directly affect the reception of the system. Experience will also affect the acceptance of the system indirectly through the trust-belief is through perception of ease of use and perception of usefulness. Mathieson [4] also suggests a relationship between experience expressed as a skill or skill with the use of technology.

Karahanna et al [16] distinguishes between adopters of information systems that have not yet been adopted with users who have adopted from time to time. Their research found that subjective norms affect the intentions of adopters of information systems that have not yet been adopted and attitudes affect adopters who have adopted over time.

This study rejected the Millati, Mudjahidin and Anggraini [10] studies which found that ease of use was more likely to be significant for users with low experience and users with high experience emphasized greater attributes of information systems (design, speed and security) than service attributes (support services customer).

From several studies it can be concluded that, factors such as perceived ease of use have different effects depending on the user experience. Inexperienced users or users with low experience, seeing ease of use is the reason for using internet banking. However in this study, the respondents' experience is quite high and has a significant relationship to intentions when moderating perceptions of ease. The lack of focus in the measurement of experience variables makes contradictory results from previous studies suggesting that low experience sees ease in determining intent. Question in the questionnaire about the experience variable in this study is less focused on asking how many times respondents use a technology or useful life in the use of technology. This results in a biased result.

### 4.4. The effect of perceived usefulness on intention to use that is moderated by the experience

Experience is a past activity in using technology to encourage the tendency to use similar technology over and over. This variable is measured through internet experience, experience in online transactions, experience using internet banking at other banks.

Perceived usefulness has no significant relationship to Intention to use that is moderated by Experience. This result is not in accordance with the initial hypothesis that there is a significant relationship between perceived usefulness to intention to use that is moderated by experience.

The process of perception depends on the past experience and the education the individual gains. The process of perception formation as the meaning of observations that begins with the stimuli. After getting stimuli, in the next stage there is a selection that interacts with interpretation, as well as interact with the closure. Selection process occurs when a person obtains information, it will take place the process of selecting messages about which messages are considered important and not important.

The results rejected the research of Millati, Mudjahidin and Anggraini and Taylor and Todd [9]. Millati, Mudjahidin and Anggraini found that users with high experience emphasized greater attributes of information systems (design, speed and security) than service attributes (customer support services) [10]. It also makes users with high experience perform the process systematically when evaluating.

Taylor and Todd in his research showed there are differences in the causes of the use of information systems by experienced users and inexperienced [9]. With experience, users have no worries about ease of use and focus on perceived benefits.

The results in this study indicate that the customer has a high experience but does not moderate the perception of the benefits to the intention in using internet banking. This inconsistency, probably due to the connection speed and security of internet banking services. Frequent slow connections make customers feel unbearable and get information optimally. Security is also an important factor in the use of internet banking. Bad experience both from personal and from the people around when transacting financial via the internet to make the user does not come back to visit the website. It can be concluded that, although the user has experience but does not moderate the perception of benefits against intent because there are several factors such as transaction security and network speed.

## 5. Conclusion

The results show that some variables have no impact on the intention of using the technology. This is because the situation in the field is different from other research. At the location of this study, the frequent problems with the internet network makes some people are not too interested in using financial transactions through the internet. Some people are more interested and feel secure when using transactions manually is coming to the place.

From several studies it can be concluded that, factors such as perceived ease of use have different effects depending on the user experience. Inexperienced users or users with low experience, seeing ease of use is the reason for using internet banking. But unlike the perceived perceive benefits, There are differences in the causes of the use of information systems by experienced and inexperienced users. With experience, users have no worries about ease of use and focus on perceived benefits.

## Acknowledgement

This research can be completed with the help of some Banks that allow researchers to examine several customers related to the intentions of the use of banking technology